\documentclass[twocolumn,showpacs,preprintnumbers,amsmath,amssymb,floatfix]{revtex4}
\usepackage{graphicx}

\begin{document}

\title{Anderson localization on Falicov-Kimball model with next-nearest-neighbor hopping and long-range correlated disorder}
\author{D. O. Maionchi$^{1}$, A. M. C. Souza$^{2}$, H. J. Herrmann$^{1,3}$, and R. N. da Costa Filho$^{1}$}
\affiliation{$^{1}$Departamento de F\'{i}sica, Universidade Federal
do Cear\'{a}, 60451-970 Fortaleza-CE, Brazil}
\affiliation{$^{2}$Departamento de Fisica, Universidade Federal de
Sergipe, 49100-000 Sao Cristovao-SE, Brazil}
\affiliation{$^{3}$Computational Physics, IfB, ETH H\"{o}nggerberg,
HIF E 12, CH-8093 Z\"{u}rich, Switzerland}
\date{\today}

\begin{abstract}
The phase diagram of correlated, disordered electron systems is
calculated within dynamical mean-field theory for the
Anderson-Falicov-Kimball model with nearest-neighbors and
next-nearest-neighbors hopping. The half-filled band is analyzed in
terms of the chemical potential of the system using the geometric
and arithmetic averages. We also introduce the on-site energies
exhibiting a long-range correlated disorder, which generates a
system with similar characteristics as the one created by a random
independent variable distribution. A decrease in the correlated
disorder reduces the extended phase.
\end{abstract}

\pacs{71.10.Fd, 71.27.+a, 71.30.+h}

\maketitle

\section{Introduction}

Many recent works \cite{cor2,cor3,cor4} have considered a
nonperturbative framework to investigate the Mott metal-insulator
transition (MIT) \cite{Mott} in lattice electrons with local
interaction and disorder using the dynamical mean-field theory
(DMFT) \cite{pap4}. For example, the Anderson transition
\cite{cor1,Anderson} has been explored on the Bethe lattice
\cite{pap1} considering the Hubbard and Falikov-Kimball models. In
these studies, the Mott MIT is characterized by opening a gap in the
density of states at the Fermi level. At the Anderson localization,
the character of the spectrum at the Fermi level changes from a
continuous to a dense discrete one.

The study of disordered systems requires the use of probability
distribution functions (PDFs). One is usually interested in typical values of these quantities
which are mathematically given by the most probable value of the
PDF. The metal and the insulator phases could be detected by
analyzing the local density of states (LDOS). In particular, the
arithmetic mean of this random one-particle quantity is noncritical
at the Anderson transition and hence cannot help to detect the
localization transition. By contrast, the geometric mean gives a
better approximation of the averaged value of the LDOS
\cite{pap1,pap2}, as it vanishes at a critical strength of the
disorder and hence provides an explicit criterion for Anderson
localization \cite{Mon,Mon2,pap3,pap4}. Recently, we adopted the
H\"older mean to analyze how the averaged LDOS depends on each
H\"older parameter that is used. We showed that the averaged LDOS
can vanish in the band center at a critical strength of the disorder
for a wide variety of averages \cite{nos}.

Most of these studies are restricted to the hopping between
nearest-neighbors (NNs). Recent works suggest that the inclusion of
next-nearest-neighbors (NNNs) hopping in the model favors the
long-range magnetic ordering \cite{eckstein}. The investigation of
these effect is particularly interesting. The breaking of
particle-hole symmetry, even at half filling, is a generic property
of real materials and creates effects on the paramagnetic phase
\cite{hirsch}, besides the frustration of the antiferromagnetic
phase \cite{anti}. The effect of nonrandom NNN hopping becomes
evident already in the noninteracting system through an asymmetric
density of states (DOS), as already derived for an arbitrary hopping
on the Bethe lattice \cite{Bethe}.

On the other hand, the long-range correlated disorder can be
generated in a variety of stochastic processes in nature \cite{pac}.
A tight-binding one-dimensional model of electronic states with the
on-site energies exhibiting long-range correlated disorder and
nonrandom hopping amplitudes was studied in ref. \cite{fran}. The
presence of an Anderson-like metal insulator transition was revealed
for a finite range of energy values where the Lyapunov coefficient
vanishes. The correlation in the disorder favors the emergence of the extended phase.

In this paper, we investigate two different aspects for the
Anderson-Falicov-Kimball model. First, considering nearest-neighbor
(NNs) and next-nearest-neighbor (NNNs) hoppings, we analyzed how
the presence of the NNN hoppings influence the phase diagram of the
ground state of this model using geometric and arithmetic averages.
Besides that, we showed how the chemical potential varies as we look
at the half-filled band of different systems; the inclusion of the
NNN hopping dislocates the half-filled band from the band center,
independent of the disorder that is considered. The ground-state
phase diagram is also presented for different values of NNN
hoppings.

Secondly, we studied the main effects of the long-range correlated disorder
(characterized by the exponent $\alpha$) in
the Anderson-Falicov-Kimball model. Unlike the one-dimensional
Anderson model, this disorder does not have a great influence on
the behavior of the system. We present the ground-state phase
diagram for electrons in a half-filled band for different values of
this disorder and the dependence of the disorder in terms of the
exponent $\alpha$ for a system without Coulomb repulsion.

In the pure half-filled Falicov-Kimball model, the Fermi energy for
electrons is inside of the correlation (Mott) gap opened by
increasing the interaction \cite{pap2}. We want to examine
how the disorder influences this gap.

The pure Falicov-Kimball model describes two species of particles,
mobile and immobile, which interact with each other when both are on
the same lattice site. This model is the simplest model to study
metal-insulator transitions in mixed valence compounds of rare earth
and transition metal oxides, ordering in mixed valence systems,
order-disorder transitions in binary alloys, itinerant magnetism
\cite{Andre}, crystallization, electronic ferroelectricity in
mixed-valence compounds, and phase diagrams of metal ammonia
solutions \cite{Leu-Csa}. It also captures some aspects of the
Mott-Hubbard MIT \cite{nos}.

The Anderson-Falicov-Kimball model considers mobile
particles that are disturbed by a local random potential, giving rise to
a competition between interaction and disorder. The Hamiltonian is
written as
\begin{equation}
H=-\sum_{ij} t_{ij}
c_{i}^{+}c_{j}+\sum_{i}\epsilon_{i}c_{i}^{+}c_{i}
+U\sum_{i}f_{i}^{+}f_{i}c_{i}^{+}c_{i},\label{Hamil}
\end{equation}
where $c_{i}^{+}$ ($c_{i}$) and $f_{i}^{+}$ ($f_{i}$) are,
respectively, the creation (annihilation) operators for the mobile
and immobile fermions (electrons and ions, respectively) at a
lattice site $i$, $t_{ij}$ is the electron transfer integral
connecting sites $i$ and $j$, and $U$ is the Coulomb repulsion
that operates when one ion and one electron occupy the same site. We
assume
$$
t_{ij} = \left\{ \begin{array}{rl}
 t_{1} &\mbox{ for nearest-neighbor sites} \\
  t_{2} &\mbox{for next-nearest-neighbor sites} \\
  0 &\mbox{ otherwise}
       \end{array} \right.
$$
The average number of electrons and ions on site $i$ are denoted,
respectively, as $n_{e}=c_{i}^{+}c_{i}$ and $n_{f}=f_{i}^{+}f_{i}$.
We consider that the occupation $n_{f}$ on the $i$th site has
probability $p$ ($0<p<1$). It was assumed, for simplicity,  that
just mobile particles are subjected to the structural disorder
\cite{pap2}. The energy $\epsilon_{i}$ is a random, independent
variable, describing the local disorder disturbing the motion of
electrons. The model is solved within the DMFT framework.

In section \ref{dmftmodel} we present the DMFT approach applied to
the Anderson-Falikov-Kimball model \cite{pap2}. In section
\ref{results1} we present the numerical results concerning the
ground-state phase diagram for the next-nearest-neighbor hopping
case. In section \ref{results2} we discuss the above model for the
long-range correlated disorder. Finally in section \ref{conclusions}
we present our conclusions.

\section{Dynamical Mean-Field Theory} \label{dmftmodel}

The formalism is based on the introduction of the hybridization
function $\eta(\omega)$, which is a dynamical mean field describing the coupling of a selected lattice site with the
rest of the system \cite{pap3}. The DMFT is calculated from the
Hilbert transforms
\begin{equation}
G(\omega)=\int \frac{ d\epsilon
N_{0}(\epsilon)}{\eta(\omega)-\epsilon+ 1/G(\omega)}, \label{g0}
\end{equation}
and
\begin{equation}
G(\omega)= \int d\omega'\frac{\rho_{q}(\omega')}{\omega-\omega'}.
\label{eq4}
\end{equation}
where $N_{0}(\epsilon)$ is the non-interacting density of states and
$G(\omega)$ the translationally invariant Green function.

The $\epsilon_{i}$-dependent LDOS is written as
\begin{equation}
\rho(\omega,\epsilon_{i})=-\frac{1}{\pi}~Im~
G(\omega,\epsilon_{i}),\label{eq1}
\end{equation}
where $G(\omega,\epsilon_{i})$ is the local $\epsilon_{i}$-dependent
Green function \cite{pap2}. $\epsilon_{i}$ is considered an
independent random variable characterized by a probability function
$P(\epsilon_{i})=\Phi(\Delta/2-\epsilon_{i})/\Delta$, with $\Phi$ being
the step function. The parameter $\Delta$ is a measure for the
disorder strength. From the $\epsilon_{i}$-dependent LDOS, we
introduce the $q$-H\"older averaged LDOS
\begin{equation}
\rho_{q}(\omega)=\left\{\sum_{i}
[\rho(\omega,\epsilon_{i})]^{q}\right\}^{1/q}\label{eq3}
\end{equation}
where the subscript $q$ defines the generalized mean. Special cases are,
for example, the minimum ($q\to -\infty$),
the geometric mean ($q \to 0$), the arithmetic mean ($q=1$) and the
maximum ($q \to \infty$). Eq. (\ref{eq3}) inserted in Eq. (\ref{eq4}) closes
the self-consistent DMFT.

\begin{figure}[t]
\includegraphics[width=90mm]{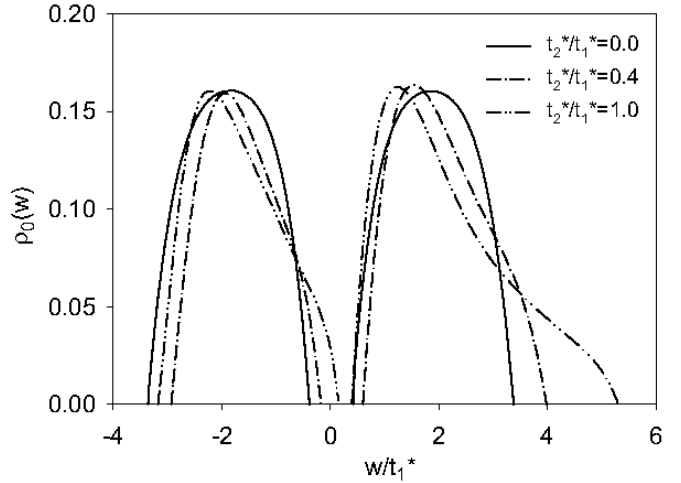}
\caption{Averaged local density of states at $U=3.6$ and $q=0$ for
(a) $t_{2}^{*}/t_{1}^{*}=0$, $0.4$  and $1.0$ for disorder strength
$\Delta=2.0$.}\label{graf1}
\end{figure}

A chemical potential $\mu$ is introduced only for the mobile
subsystem, to fix the system in the half-filled band
($n_{e}=p=1/2$). For the Bethe lattice in the limit of infinite
connectivity $K$, one can use the scaling
$t_{1}=t_{1}^{*}/\sqrt{K}$ and $t_{2}=t_{2}^{*}/K$. In the
nearest-neighbor hopping case ($t_{2}=0$) the density of states is
the semi-elliptic function
$N_{0}(\epsilon)=\sqrt{4-(\epsilon/t_{1}^{*})^2}/(2\pi t_{1}^{*})$
\cite{Bethe}, where the bandwidth is $4t_{1}^{*}$. We can find
$\eta(\omega)=t_{1}^{*2} G(\omega)$ \cite{pap2}. In this case, the
ground-state properties in the half-filled band case are solely
determined by the quantum states in the band center ($\omega=0$),
and we can determine the transition points in the phase diagram by
linearizing the DMFT equations. However, for $t_{2} \ne 0$ the
symmetry in the band center is absent, and we can not use a
recursive relation within the linearized DMFT.

For the next-nearest-neighbor hopping case ($t_{2} \ne 0$) the
analytical expression for the density of states on the Bethe lattice is
given by
\begin{equation}
N_{0}(\epsilon)= \Theta (t_{1}^{*2}+4t_{2}^{*2}+4t_{2}^{*}\epsilon)
\frac{ \sqrt{4-\lambda_{+}^2} + \sqrt{4-\lambda_{-}^2}} {2\pi
\sqrt{t_{1}^{*2}+4t_{2}^{*2}+4t_{2}^{*}\epsilon} } \label{n0}
\end{equation}
where $\Theta(\epsilon)$ is the step function and $\lambda_{\pm
}(\epsilon)$ is solution of the equation
\begin{equation}
t_{2}^{*} \lambda^{2} + t_{1}^{*}\lambda - (t_{2}^{*}+\epsilon) = 0.
\end{equation}

\begin{figure}[t]
\includegraphics[width=90mm]{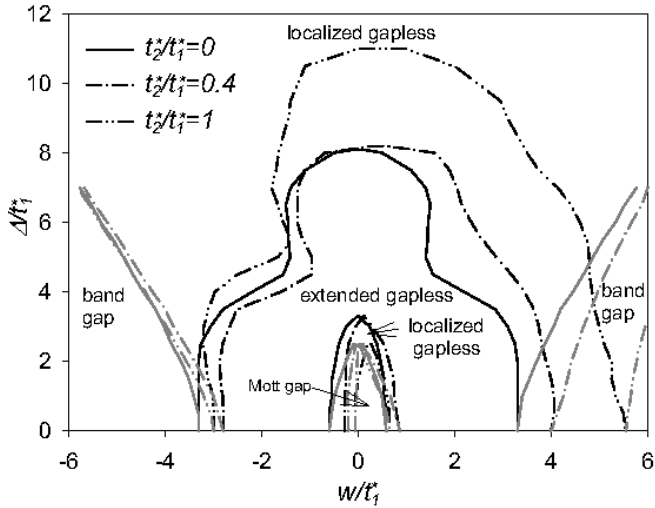}
\caption{Phase diagram of the ground state for the
Anderson-Falikov-Kimball model for $U=3.6$ at $t_{2}=0$ (solid
line), $t_{2}^{*}/t_{1}^{*}=0.4$(dash-dot line) and
$t_{2}^{*}/t_{1}^{*}=1$ (dash-dot-dot line). Black lines present
mobility edges determined within DMFT with geometric averaging
($q=0$) and brown lines show band edges determined within DMFT with
arithmetic averaging ($q=1$).} \label{spectral}
\end{figure}

\begin{figure}[b]
\includegraphics[width=90mm]{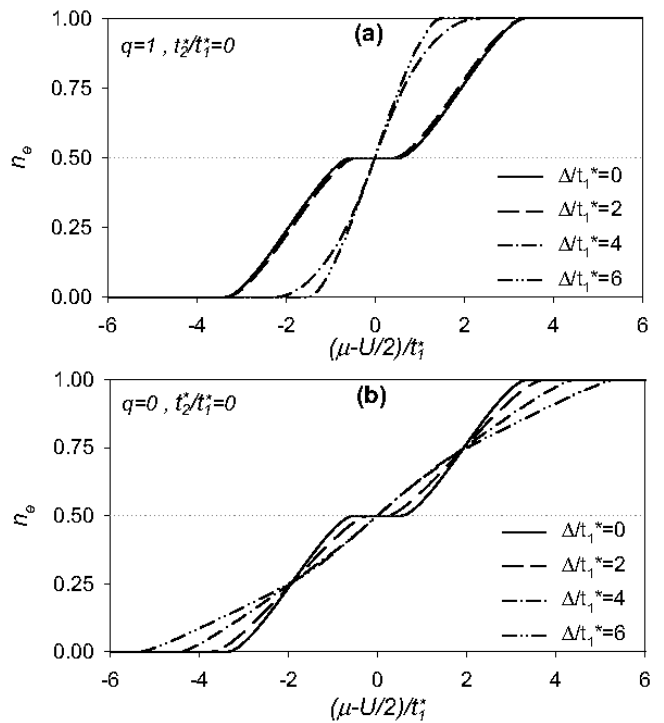}
\caption{Averaged electron number $n_{e}$ versus chemical potential
at $U=3.6$, for $t_{2}^{*}/t_{1}^{*}=0$ and $\Delta = 0$, $2.0$,
$4.0$ and $6.0$. (a)$q=1$ and (b)$q=0$.}\label{var1}
\end{figure}

\section{Next-Nearest-Neighbor Hopping}\label{results1}

We considered that the initial value of $\rho(\omega,\epsilon_{i})$
is a uniform distribution with bandwidth $W=8t_{1}^{*}$, then we
determined $G(\omega)$ in order to obtain $\eta(\omega)$ and finally
the new values of $\rho(\omega,\epsilon_{i})$. This loop is
performed until we find the stable configuration for
$\rho_{q}(\omega)$. The relation between $G(\omega)$ and
$\eta(\omega)$ is obtained in a straightforward way from Eqs.
(\ref{g0}) and (\ref{n0}) leading to the result
\begin{equation}
\eta(\omega) = G(\omega) \left\{
\frac{t_{2}^{*2}}{1-G(\omega)t_{2}^{*}} +
\frac{t_{1}^{*2}(1-G(\omega)t_{2}^{*})}{(1-2G(\omega)t_{2}^{*})^{2}}
\right\}.
\end{equation}

If we consider fixed values of $U$ and $t_{2}^{*}/t_{1}^{*}$ we can
observe the dependence of the averaged LDOS on the value of $\Delta$
that is used. Some of these results are presented in Fig.
\ref{graf1}. We consider the next-nearest-neighbor hopping model
($t_{2}^{*}/t_{1}^{*}=0.0$, $0.4$ and $1.0$). We can see that the
behavior of the LDOS is not the same in all cases. In contrast to
the case $t_{2}^{*}/t_{1}^{*}=0$, if $t_{2}^{*}/t_{1}^{*} \ne 0$ the
$\rho_{q}(\omega)$ is asymmetric.

A signature of the Anderson localization is the vanishing of
$\rho_{q}(\omega)$ as we increase $\Delta$. The detection of the
Anderston localization depends on the average that is used to
calculate $\rho_{q}(\omega)$; using the geometrical average ($q=0$),
$\rho_{q}(\omega)$ vanishes for a certain value of $\Delta$, while
using the arithmetical average ($q=1$), the localization cannot be
detected. Fig. \ref{spectral} presents the phase diagram of the
ground state for the Anderson-Falikov-Kimball model for $U=3.6$ at
$t_{2}=0$ (solid line), $t_{2}=0.4$(dash-dot line) and $t_{2}=1$
(dash-dot-dot line) using the geometrical average (black lines) and
the arithmetic average (brown lines). It is obtained observing the
behavior of the average local density of states for different values
of $w$ as the disorder strength $\Delta$ is increased. We can
identify three regions with respect to the states at the Fermi
level: extended gapless phase (continuous spectrum), localized
gapless phase (pure point spectrum) and gap phase \cite{pap2}. The
lines delimiting these phases are centered around $w=0$ for
$t_{2}^{*}/t_{1}^{*}=0$, what does not happen for
$t_{2}^{*}/t_{1}^{*} \ne 0$. The asymmetry in the phase diagram
becomes more evident, the higher the value of $t_{2}^{*}/t_{1}^{*}$.

\begin{figure}[t]
\includegraphics[width=90mm]{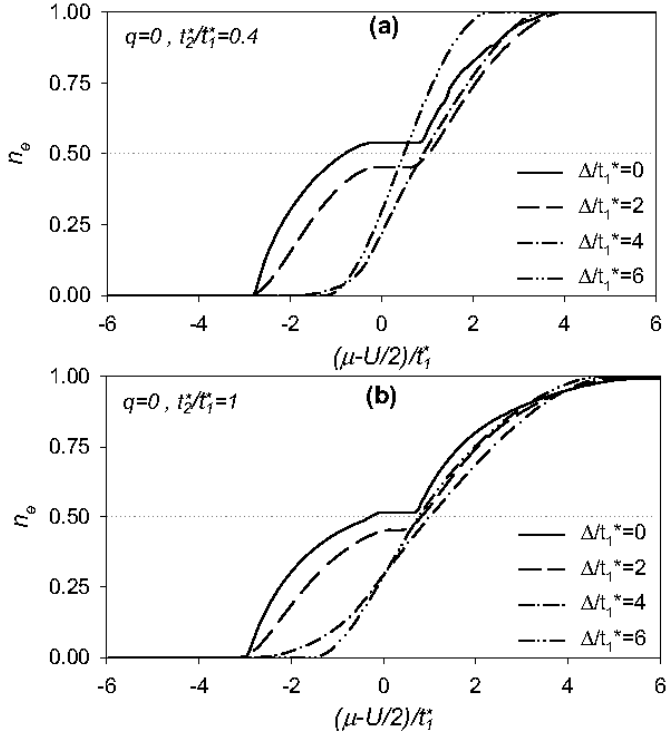}
\caption{Averaged electron number $n_{e}$ versus chemical potential
at $U=3.6$ using $q=0$, for $t_{2}^{*}/t_{1}^{*} \ne 0$ and
$\Delta=0$, $2.0$, $4.0$ and $6.0$. (a) $t_{2}^{*}/t_{1}^{*}=0.4$
and (b) $t_{2}^{*}/t_{1}^{*}=1.0$.}\label{var2}
\end{figure}

\begin{figure}[t]
\includegraphics[width=90mm]{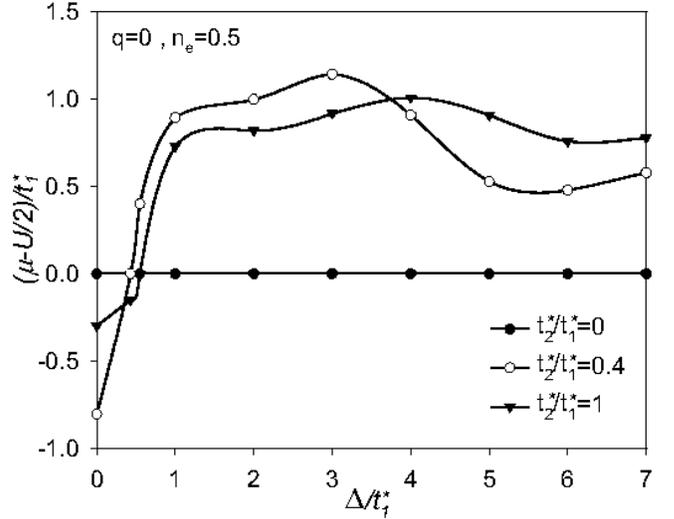}
\caption{Chemical potential versus disorder parameter $\Delta$ for
the half-filled band ($n_{e}=0.5$). The results are obtained for
$t_{2}^{*}/t_{1}^{*}=0$, $0.4$ and $1.0$.}\label{ult}
\end{figure}

As already pointed out, if $t_{2} \ne 0$, the chemical potential $\mu =
U/2$ that was introduced does not affect the system in the
half-filled band. When we look at the curves of Fig. \ref{var1}
obtained for the relation between the chemical potential $\mu$ and
the average number of electrons $n$, we see that all the curves
obtained for $t_{2}^{*}/t_{1}^{*}=0$ are centered at $\mu = U/2$ and
$n_{e}=0.5$, independent of the average that is considered. This
means that the half-filled band of the system is always in a band
center ($\omega=0$) as we vary the disorder $\Delta$ of the system.
In Fig. \ref{var2} we present the curves corresponding to $t_{2} \ne
0$ ($t_{2} = 0.4$ and $t_{2} = 1.0$) using the geometrical average.
We can observe in all curves that for a half-filled band system
($n_{e}=0.5$) the value of $\mu$ does not correspond to $U/2$
anymore, independent of the average that is used and the disorder
that is imposed.

\begin{figure}[b]
\includegraphics[width=90mm]{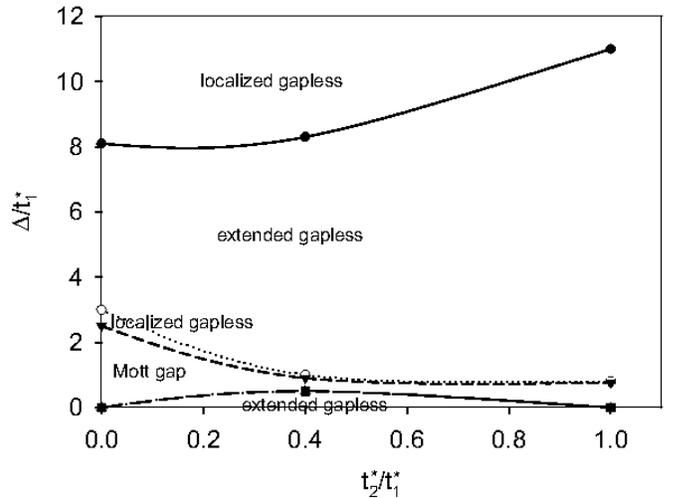}
\caption{Ground-state phase diagram for electrons in a half-filled
band as a function of $t_{2}^{*}/t_{1}^{*}$ for $U=3.6$ . Dots are
determined from the numerical solution of the DMFT equations. Lines
are guide to the eye.} \label{pd1}
\end{figure}

\begin{figure}[t]
\includegraphics[width=90mm]{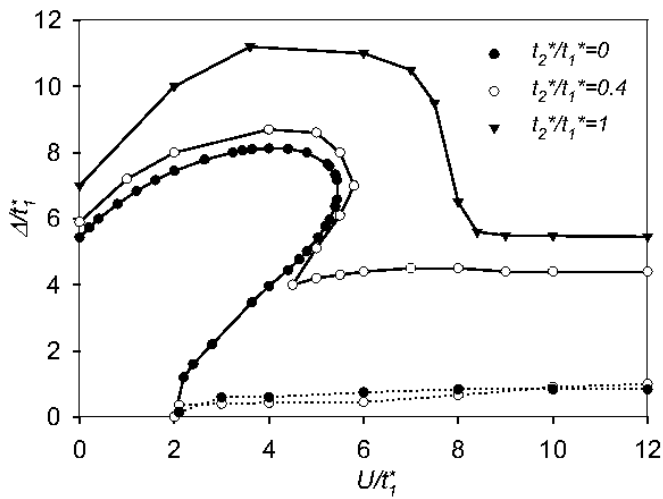}
\caption{Ground-state phase diagram for electrons in a half-filled
band as a function of $U$ for $t_{2}^{*}/t_{1}^{*}=0$, $0.4$ and
$1.0$. Dots are determined from the numerical solution of the DMFT
equations. Lines are guide to the eye.}\label{cor1}
\end{figure}

Fig. \ref{ult} shows the chemical potential versus disorder
parameter $\Delta$ for the half-filled band ($n_{e}=0.5$). We observe
that for $t_{2}^{*}/t_{1}^{*}=0$ the half-filled band is in a band
center ($\omega=0$) independent of disorder $\Delta$. As we include
the influence of next-nearest-neighbors $t_{2}^{*}/t_{1}^{*} \ne 0$
the half-filled band is dislocated to bands with $\omega > 0$ that
correspond to values of $\mu$ always greater than $U/2$.

The ground-state phase diagram for electrons in a half-filled band
as a function of $t_{2}^{*}/t_{1}^{*}$ for $U=3.6$ is shown in Fig.
\ref{pd1}. Performing the same process for other values of $U$, we
can easily obtain all the simulation points of Fig. \ref{cor1}. In
this figure we present a complete ground-state phase diagram for
electrons in a half-filled band for three different values of for
$t_{2}^{*}/t_{1}^{*}$, namely $t_{2}^{*}/t_{1}^{*}=0$, $0.4$ and
$1.0$. The presence of $t_2$ increases the extended phase. The dashed
line represents the values of $\Delta$ and $U$ where the energy gap
goes through the half-filled band. Note that for $t_2$ different from
zero and $U$ large the extended phase does not disappear because the
energy gap does not correspond to the half-filled band.

\section{Long-Range Correlated Disorder}\label{results2}

Recently results \cite{fran} have shown that localization
properties are modified if correlations are introduced in the
disorder distribution. For example, long-range correlated disorder
favors the delocalization, giving in 1D systems a range of energies
with extended eigenstates. The one-dimensional Anderson model
with long-range correlated diagonal disorder displays a phase of
extended electronic states where the on-site energy disorder
distribution is described by a power-law spectral density.

\begin{figure}[b]
\includegraphics[width=80mm]{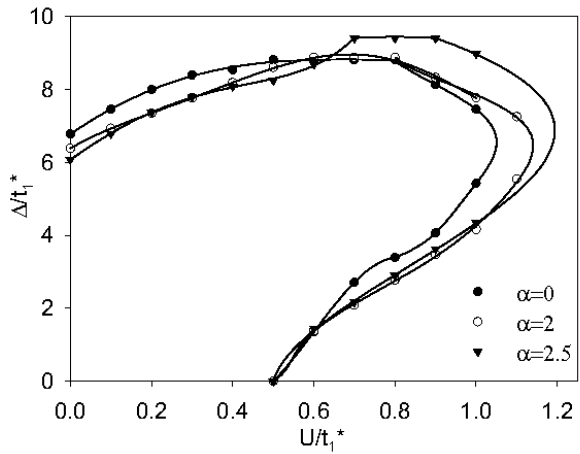}
\caption{Ground-state phase diagram for electrons in a half-filled
band by using long-range correlated disorder for $\alpha=0$, $2.0$
and $2.5$. Dots are determined from the numerical solution of the
DMFT equations. Lines are guide to the eye.}\label{cor2}
\end{figure}

\begin{figure}[t]
\includegraphics[width=80mm]{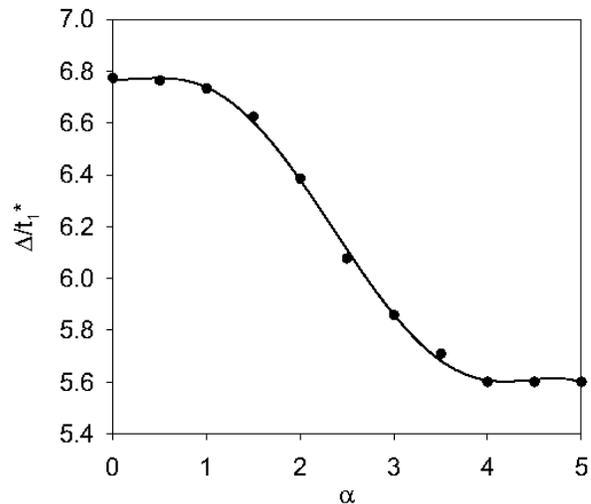}
\caption{Ground-state phase diagram for electrons in a half-filled
band versus $\alpha$ for $U/t_{1}^{*}=0$. Dots are determined from
the  numerical solution of the DMFT equations. Lines are guide to
the eye.}\label{cur}
\end{figure}

In what follows, we will study the effect due to long-range
correlated disorder in the Anderson-Falicov-Kimball model on the
Bethe lattice using the DMFT approach. The energy $\epsilon_{i}$
will not be considered as a random independent variable. Following
an approach based on discrete Fourier transforms to
construct long-range correlated sequences, the on-site energies can
be generated by the expression \cite{fran}
\begin{equation}
\epsilon_{i} = \sum_{k=1}^{N/2} k^{-\alpha/2} \left\|
\frac{2\pi}{N}\right\|^{(1-\alpha)/2} \cos{ \left\{ \frac{2\pi
ik}{N}+\phi_{k} \right\} },
\end{equation}
where $N$ is the number of sites and $\phi_{k}$ are $N/2$
independent random phases uniformly distributed in the interval
[0,2$\pi$]. The uncorrelated disorder is recovered for $\alpha=0$.

We will normalize the energy sequence to have the mean value
$<\epsilon_i> = 0$. The standard deviation is defined as $\sigma =
\sqrt{<\epsilon_i^2> - <\epsilon_i>^2} = 6 \Delta /\sqrt{3}$. For $\alpha
= 0$, we obtain a gaussian distribution for the $\epsilon_i$. In
this sense, the values of the Coulomb repulsion $U$ and the
disorder $\Delta$ that we obtain when $\rho_0(w) = 0$ are not the
same as obtained with the uniform distribution.

In the case $U = 0$, we find in the literature that for the cubic
network the transition from the extended states to the localized
ones occurs for $\Delta_c = 5.5$ for the uniform distribution and
$\Delta_c = 7.0$ for the gaussian distribution \cite{rd,ult}. For
the same value of $U$, we found in our simulations $\Delta_c = 5.7$
and $6.8$, respectively, for the uniform and the gaussian
distributions.

In Fig. \ref{cor2}, we present the ground-state phase diagram for
electrons in a half-filled band by using long-range correlated
disorder and the geometric average. We see that for different values
of $\alpha$, the diagram that is obtained is similar to the one that
corresponds to uncorrelated disorder.

The difference between the diagrams can be quantified for different
values of the exponent $\alpha$ when we look at the value of
$\Delta$ when $U=0$. The result is presented in Fig \ref{cur}. We
see that the value of $\Delta$ diminishes as the value of $\alpha$
is increased and tends to a constant value for $\alpha > 4$. The
decrease of $\Delta$ means that the correlation reduces the extended
phase, different to what happens in low dimension \cite{fran}. 

\section{Conclusions}\label{conclusions}

In the present paper, we studied the solutions of the
Anderson-Falicov-Kimball model with nearest-neighbor (NNs) and
next-nearest-neighbor (NNN) hopping. We found the averaged LDOS
calculated using the arithmetic and geometric mean within dynamical
mean field theory. We showed that the inclusion of the NNN with
$t_{2}^{*}/t_{1}^{*} \ne 0$ moves the half-filled band from the
band center ($\omega=0$) to bands with $\omega > 0$, independent of
the disorder $\Delta$ that is considered. Besides that the chemical
potential $\mu$ changes from $U/2$ for $t_{2}^{*}/t_{1}^{*} = 0$ to
greater values for $t_{2}^{*}/t_{1}^{*} \ne 0$.

We also analyzed the behavior of the system when long-range
correlated disorder is considered using both the geometric and
arithmetic means. We showed that the inclusion of this kind of
disorder does not present results different from the ones already
obtained for an uncorrelated disorder system \cite{nos}. At the
other hand, we showed that decreasing the correlated disorder
reduces the extended phase for $U=0$, different to what happens in
low dimension.

A previous work \cite{eckstein} about phase separation in the
particle-hole assymetric Hubbard model involving NNN hoppings
suggests that the inclusion of the temperature in determining the
$(U,T,\mu)$ phase diagram is crucial for the study of the Anderson
localization in a disordered electron system described by the
Falicov-Kimball model. We will analyze these effects in future works.

\section{Acknowledgments}
Financial support of Conselho Nacional de Pesquisas Cient\'ificas
(CNPq) is acknowledged. HJH thanks the Max Planck prize.
AMCS thanks the International Centre for
Theoretical Physics (Trieste, Italy) for the support and hospitality
during his visit from January to March 2008 as ICTP Associate.


\end{document}